\documentstyle[psfig,12pt,aaspp4]{article} 

\begin{document}

\title{X-ray variability of GRS 1915+105
during the low-hard state observed with the
Indian X-ray astronomy experiment (IXAE)}

\author{B. Paul, P. C. Agrawal, A. R. Rao, M. N. Vahia, and J. S. Yadav}
\affil{Tata Institute of Fundamental Research, Homi Bhabha Road, Mumbai 400 005, India}
\affil{e-mail: bpaul@tifrvax.tifr.res.in (BP), pagrawal@tifrvax.tifr.res.in (PCA),
arrao@tifrvax.tifr.res.in (ARR), vahia@tifrvax.tifr.res.in (MNV) and jsyadav@tifrvax.tifr.res.in (JSY)}
	\and
\author{T. M. K. Marar, S. Seetha and K. Kasturirangan}
\affil{ISRO Satellite Centre, Airport Road, Vimanpura P.O. Bangalore  560 017, India.}
\affil{e-mail: seetha@isac.ernet.in (SS)}

\begin{abstract}

The galactic superluminal transient X-ray source GRS 1915+105 was
observed with the pointed proportional counters (PPCs) onboard the
Indian satellite IRS-P3 during 1996 July 23-27. We report here
details of the behavior of this source during the relatively quiet
and low luminosity state. Large intensity variations by a factor
of 2 to 3, generally seen in black-hole candidates, are observed
at a time scale of 100 ms to few seconds. No significant variation
is detected over larger time scale of minute or more. The intensity
variations are described as sum of shots in the light curve, and
the number distribution of the shots are found to be exponential
function of the fluence and duration of the shots. The cross
correlation spectrum between 6-18 keV and 2-6 keV X-rays is
found to have asymmetry signifying a delay of the hard X-rays by
about 0.2 to 0.4 sec. This supports the idea of hard X-rays being
generated by Compton up-scattering from high energy clouds near
the source of soft X-rays. Very strong and narrow quasi periodic
oscillations in the frequency range 0.62 to 0.82 Hz are observed.
We discuss about a model which explains a gradual change in the
QPO frequencies with corresponding changes in the mass accretion
rate of the disk.

\end{abstract}

\section{Introduction}

The X-ray transient source GRS 1915+105 was discovered in 1992
in hard X-rays using the WATCH all sky X-ray monitor onboard the GRANAT
satellite
(\cite{cast:94}). Large intensity variations in the source over
time scales of a few hours to a few days were detected. During two
years of hard X-ray observations by WATCH, two powerful bursts
were discovered during which the source luminosity was as high
as $10^{39}$ ergs s$^{-1}$. The hard X-ray spectrum was found to
fit well with a power law spectrum with a photon index $(\alpha)$
of -2.5 (\cite{sazo:94}).

A probable optical counterpart of the source has been observed
only in the I band at 23.4 magnitude (\cite{boer:96}). The
optical faintness is probably owing to a low mass companion
and very high extinction at a distance of 12.5 kpc. H and He
emission lines in the infrared were found to be narrow and no
Doppler shift was observed. Absence of high velocity signatures
in the line profiles in the IR band suggests that the companion
is a low mass star (\cite{cast:96}). However, high resolution
spectral observations in the near-infrared revealed similarity
with spectra of very high mass stars of Oe or Be spectral
type (\cite{mira:97}).

Superluminal motion of two symmetric radio emitting jets of GRS
1915+105 was discovered by Mirabel \& Rodriguez (\cite{mira:94})
which prompted the object being termed as a micro-quasar.
Correlated enhanced radio
and hard X-ray emissions were discovered from the source in a long
time monitoring during 1994 September to 1996 March (\cite{fost:96}).
In a previous outburst during 1993 December to 1994 April decreases
or dips in the hard X-ray flux were observed during the radio
flares. This observation suggested an interaction between the hard
X-ray emission and the jet production. Redirection of the
accreted material onto jets can cause the observed X-ray and radio
intensity fluctuations (\cite{harm:97}). Radio flares are thought
to be synchrotron emission of outgoing plasmoids from the central
object. Infrared jets at the same position angle as that of the radio jets
were discovered by Sams et al. (\cite{sams:96}). The I-R emission
from this source is described to be {\it free-free} emission of a
wind flowing out of the disk. Correlation between the length of a jet,
its brightness temperature and central source luminosity was observed
which is similar to that seen in AGNs and quasars. This opened up the
possibility of understanding the processes in the centers of the AGNs
by studying GRS 1915+105 for which evolution time scale of the jets is a
much shorter one.

The X-ray luminosity increased to a very high level in 1996 and
the source was observed on several occasions by the pointed
proportional counters (PPCs) onboard the Indian satellite IRS-P3
(\cite{agra:96a}) and also by PCA and ASM detectors on board
the RXTE (\cite {brad:96}). From 1996 February to March 28 the intensity
varied erratically around a mean intensity level of about 1 Crab and
this state is termed as 'chaotic state'. During March 28 to May 17
the source was in a very high intensity state in a declining phase
but the intensity variations were relatively less and this is named
'bright state' of the source. During May 17 to July 16 very large
intensity flares were observed. In this period the X-ray
intensity is found to increase by a factor as large as 6 with time
scale of a few days. GRS 1915+105 was found in the same state again
during August 15 to December 30, and this state is called 'flaring
state'. In between these two flaring states i.e, during July 16 to
August 15 and after the end of the second flaring state until the
present time ASM observations indicate a stable state of the source
with luminosity about 500 mCrab in the 1.5-12 keV observation range
of ASM. The relatively quiet state of the source is termed 'low-hard'
state because the spectrum was harder during this low state which is
common in black hole candidates.

Strong, narrow, quasi periodic oscillations of varying frequency were
discovered in GRS 1915+105 with the PPC observations (\cite{agra:96b};
\cite{paul:97}).
Intensity dependent narrow QPOs were detected with the PCA. Strong
harmonics were also seen at lower frequencies which vanished with
increase in intensity and QPO frequency (\cite{chen:97}). The timing
behavior, is on some occasions similar to that of the black-hole candidates
GS 1124-68 and GX 339-4 (\cite{bell:97}). The X-ray intensity was
found to vary on a variety of time scales ranging from sub-seconds to days
during the flaring state, and the spectrum also changed
during the brightness variations (\cite{grei:96}).
In the color intensity diagram two separate branches are clearly observed.
According to the intensity states, timing behavior and spectral
informations, the source has four different states none of which is
typical of a black hole state (\cite{morg:97}).

During our observations in the 2-18 keV energy band spanning 5 days the
source was in the low-hard state. The power density spectrum (PDS) of
GRS 1915+105 during the low-hard state has a
broken power law shape which is almost flat below 1 Hz and
decreases sharply above the QPO frequency (\cite{paul:97}).

\section{Instrument and Observations}

The Indian X-ray Astronomy Experiment (IXAE) consists of three
pointed proportional counters (PPCs) and one X-ray sky monitor
(XSM) (see \cite{agra:97} for more details). The Indian satellite
IRS-P3 carrying these instruments along with two remote sensing
payloads, was launched with the Polar Satellite Launch Vehicle
(PSLV) from Shriharikota (India), on 1996, March 21. The satellite
orbits at an altitude of 830 km with an inclination of $98^\circ$ 
with the equator. Observations of X-ray sources are carried out
only in those parts of the orbits which do not go through
the South Atlantic
Anomaly (SAA) region. The satellite orbit being nearly polar the
charged particle background in the detectors increases near the
poles and the good observation band is defined to be within 
the latitude band of $30^\circ$ S to $50^\circ$ N.
The operating region is chosen
to be in this latitude range where the detector background
is found to remain constant.

The PPCs have an energy range of 2-18 keV with 60\% efficiency
at 6 keV. The typical energy resolution of a PPC averaged over
the entire detector is 22\% at 6 keV.
Total photon collecting area of the three
detectors is 1200 cm$^2$. The detectors gain can be controlled
by changing the high voltage supply in discrete steps and hence
the effective energy
range can be kept within the desired value. The X-rays
detected in each PPC are analyzed in  processing
electronics units (one for each detector). The processed data, which
contain the pulse height histograms and the count rate profiles, are
stored in the onboard memories. The timing
resolution is different for different modes of observations which
is driven by scientific need for the source being observed.
The spectral 
response and detection efficiency of the PPCs has been calibrated by
observing the bright X-ray source Crab nebula.

The galactic superluminal X-ray source GRS 1915+105 was observed
during 1996 July, 23-27 with the PPCs for a total duration of 8850
sec.  Observations were limited to the satellite position between 
the latitude $30^\circ$ S and $50^\circ$ N to avoid
high background regions induced by the
charged particles. All but one observations were done in a mode with
timing resolution of 100 ms, and 64 channel energy spectrum of the
source was stored every 10 sec.  Observation stretches were at
most about 20 minutes long and in some cases of smaller duration.

\section{Analysis and results}

Background observations were made during August 1-3 in a nearby region
of the sky which is devoid of any known X-ray source. When data in the
good latitude region of $30^\circ$ S to $50^\circ$ N are taken, the
background count rate fits well with a constant value with reduced
$\chi^2 \sim 1.4$. Background subtracted light curve is generated for
each of the observation slots and the total light curve is shown in
Fig. 1. Individual light curve of each observation with integration time
of 1 minute are plotted in Fig. 1. The individual observation stretches
are 2 to 19 minutes long. The date and time of the observations
are shown in the panels. The bottom panel shows the ASM light curve of
GRS 1915+105 in 1.5-12 keV range during 1996 July 20-29. Each data
point is a result of about 90 seconds observation of ASM with 5-10
observations every day. The ASM light curve of the source, as shown 
in the lower panel of Fig. 1., is also featureless during the days of
our observations. The intensity decrease as observed on July 27 is not
evident in ASM data, but the PPC and ASM observations are not
exactly simultaneous. Day to day variability in the source intensity
is within 10\% of the average value except for the final day of
observation. The rms variability in the 1 minute light curve of
individual observation slots is only about 1.6\%, a part of which is
also due to the nonstatistical variation in the background. This small
variation can be compared to the 0.5\% rms variation estimated for a 
constant intensity light curve of the same intensity with only Poissonian
variations.

The hardness ratio, defined as the ratio of counts rate in the 5-12
keV band and 3-5 keV band is
found to decrease gradually with few days time scale in the ASM
observations during the soft-hard state. We observe no noticeable
change in the hardness ratio during individual
observation slots. In Fig. 2. we have shown a section of the light
curve with the hardness ratio plotted with it. The hardness ratio
is defined as the number of counts in 6-18 keV divided by
those in 2-6 keV. A bin size of 5 seconds has been used to compute the
hardness ratio to reduce the error bars.

A search was made to find intensity variations in the source largely
exceeding the photon counting statistics. Each individual time bin
was inspected with respect to a running average in the light curve
around that bin and intensity variations above the average were
classified in terms of its strength. In Fig. 3a. we have shown the
number distribution of data bins exceeding the average, as a function
of the excess. The number of data points where large intensity
enhancement is detected is much more than what expected in an
otherwise constant intensity light curve with Poissonian statistics.
This difference is more pronounced for the larger intensity
enhancements.  We have earlier reported large intensity variations
over time scale of 100 ms to few sec in our observation of GRS
1915+105 (\cite{paul:97}). However there is no intensity variation
at a longer time scale of a minute or more as can be seen in Fig. 1.
where one minute count rate is plotted.

Time variability in the X-ray intensity of black hole sources
has been proposed to be the result of randomly occuring shots with
exponential rise and/or decay (\cite{terr:72}). A Large number
of shots in the Cyg X-1 light curve were added and the resultant
profile was found to have nearly symmetric rise and decay
(\cite{nego:94}). To
quantify the variations in the intensity as sum of shots in the
light curve of GRS 1915+105, we have identified shots and classified
them in terms of the number of photons in them. Every data
bin of the 100 ms light curve is compared to a running average
around it, and successive data bins, when found to be above the
average, a shot is presumed to have occured. The total excess
counts in the individual shots above the average are calculated
and a number distribution of that is shown in Fig. 3b. We find that
the distribution fits very well with an exponential function
({\bf f(S) = N~exp(S/C); \it with N = 4140 and C=10.7; S is the
strength of the shot in photon counts}). The durations of
the shots are shown in Fig. 3c. which also has an exponential
form with increasing slop above 0.7 seconds. The shape
of the curve in Fig. 3c. can also be explained as an exponential
distribution of shot duration with a hump around 0.7 sec, which
is the width of the pulse profile at the quasi-periodic oscillation
period of 1.4 second.

To measure any delay between the hard and soft X-rays, cross
correlations (correlation coefficients with different delays) were
calculated. All observation slots were divided into smaller
segments of 64 data points of 100 ms duration. The cross correlation
function between the 2-6 keV and 6-18 keV count rate profiles were
calculated for all of these small data lengths. The resultant
cross correlation functions were added and averaged and are plotted
in Fig. 4. The peaks in the cross-correlation plot are due to the
strong QPOs at a frequency of 0.7 Hz. The region near 0 is plotted
in the inset and the asymmetry around 0 delay is clearly visible. One
possible explanation for this asymmetry in the cross correlation function
is a time lag between the soft and hard X-ray oscillations.
The difference between the right and left hand side of the cross
correlation function is maximum at around 0.2-0.4 sec indicating a
delay of 0.2 to 0.4 sec for the hard X-rays compared to
the soft X-rays in our observations. Similar asymmetry is observed
in all the observations and in all the three detectors.

We have discovered quasi-periodic oscillations in GRS 1915+105
with frequency varying between 0.62 to 0.82 Hz
(\cite{agra:96b}; \cite{paul:97}). GRS 1915+105 is the sixth black
hole candidate after GX 339-4, Cyg X-1, LMC X-1, GS 1124-68 and GRO J0422+32,
in which QPOs have been observed. The power density spectrum obtained from
the PPC observations shows
that the QPOs are very narrow ($< 0.2$ Hz) and strong (rms 9\%). The
PDS is flat for frequencies less than the QPO frequency and at
frequencies above this it follows a power law of index -1.5. There
is no marked difference between the power spectrum of the low and high
energy X-rays. The PDS of GRS 1915+105 as observed
in its low state resembles that of the other black hole candidates in
their very high state.
This type of band limited noise is characteristic of black hole
sources in their very high state (\cite{miya:92}; \cite{klis:95}).
Black hole sources have strong very low frequency intensity
variations when in low-hard state and in high state the
PDS is flat below a break frequency.

The light curve was folded at the observed QPO period and the
resultant profile is found to be nearly sinusoidal with pulse
amplitude of 4\% of the total intensity of the source (Fig. 5).
There is no noticeable difference in pulse shape in the two energy
ranges, 2-6 and 6-18 keV.

\section{Discussion}

Large intensity variations over very short time scales are present in many
black-hole candidates and in some neutron star systems. In the neutron star
systems if the magnetic field is weak, the inner radius of the
accretion disk can be very close to the neutron star surface resulting in
flux variation at the Keplerian frequencies, or at the
beat frequencies between the Keplerian frequency and spin of the
neutron star. In the case of black-hole sources, the disk can extend upto the
lowest stable orbit (3 times the Schwartzchild radius) for a
non-rotating black hole, or even closer to the compact object for a
maximally rotating compact object. The fast variations observed in
GRS1915+105 indicate size of the emission region $\la$ 0.1 light second,
large for the inner disk of a few stellar mass black-hole.

The hard X-rays are found to be delayed compared to the soft X-rays
indicating that the soft and hard X-ray emitting regions are not the
same and the variations in luminosity of the two regions are not
simultaneous, as it is expected to be for physically separated regions.
The hard X-ray emissions from black hole candidates are usually thought
to be reprocessed emission of underlying soft photons from an inner disk.
In such cases multiple inverse Compton scattering by the energetic
electrons in the corona, which is responsible for the increase in energy 
of the photons, may introduce some delay between the soft and hard X-ray
spectrum. Our observation of hard X-ray delay in the source, which is
overlying on the QPOs as shown in Fig. 4., indicates that the soft X-rays
(most of them of energy $<$ 6 keV) from the inner disk are up-scattered
by a plasma of size of a fraction of a light-second. 

Quasi-periodic oscillations of many different frequencies are observed
in GRS 1915+105 (\cite{morg:97}). During the bright state, low frequency
QPOs were observed with harmonics. QPOs of two or more different
frequencies were also observed. The QPO frequency in the flaring
state varied erratically between 0.0016 Hz and 0.16 Hz with a higher
frequency QPO of 7.6 Hz observed on May 14. In the two flaring
states the QPO frequency is relatively high, 1.5 Hz to 8 Hz 
varying  erratically with a low frequency QPO of 0.003 Hz observed
on June 16. The 67 Hz QPOs are observed during the bright and
flaring state. The QPOs frequency variations during the low-hard
sate, seems to have some definite pattern in it. The QPO frequency
was 2.3 Hz at the beginning of the low-hard state, decreased
slowly to a lowest frequency of 0.62 Hz on July 25, and again
increased slowly to 2.0 Hz towards the end of the low-hard
state. The QPO frequency history of GRS 1915+105 during its
low state inferred from published data is shown in Fig. 6.
The rms/mean of 6 days observations of ASM around each data
point is plotted along with the QPO frequency.
Crossed circles are PPC observations, the dotted circles represent
PCA observations (\cite{morg:97}) and the stars are the rms/mean
obtained from ASM archival data.
We notice that the QPO frequency
decreased to a minimum during our observations and
then increased again. Erratic changes in the QPO frequency between 
segments of observations on the same day are present. An overall
trend of a smooth decrease followed by increase in the QPO frequency 
is clearly observed. The rms deviation of the X-ray intensity shows
dramatic change when the source makes transition from flaring to low state and
and again back to the flaring state. The QPO frequency appears to be
very strongly related to the rms/mean intensity of the source.
The hardness ratio between the 5-15 keV flux and 3-5 keV flux
increased during the low-hard state and decreased again as the source
went to the flaring state again. 

The quasi-periodic oscillations detected during our observations are
stable, narrow and strong. It may represents the Keplerian
motion of blobs of hot material in the disk. Stability of the QPO
frequency in the range 0.62 to 0.82
Hz during 5 days of observations, during which the
overall intensity of the source is also nearly constant, indicates that
the QPO production region is narrow in the radial direction and is
stable when the mass accretion rate is stable. But the force behind
confinement of the hot spots in the radial directions which can produce
narrow QPOs, is not understood.

In a model of disk accretion around black hole sources, in which 
the disk passes through a standing shock close to the centrifugal
barrier, a QPO generation mechanism has been developed (\cite{molt:96}).
Oscillations around the mean shock location can give rise to
intensity modulations similar to the quasi periodic oscillations
observed in the black hole candidates like GX 339-4, GS 1124-68 etc.
The oscillations of the shock surface arise as a balancing act
between the infall of the converging disk material and the outflow
of the hot shockfront. If the cooling time scale of the expanding
postshock halo and the infall time scale of the disk are nearly
the same, a modulation of upto 10-15\% of the intensity can be
produced. In this model the QPO frequency is predicted to increase
with mass accretion rate. Sustained quasi-periodic oscillations
are possible if the dominant cooling process
is bremsstrahlung. Chakrabarti \& Titarchuk (\cite{chak:95}) have
shown that for sources emitting near their Eddington luminosity the
post shock region emits mainly by bremsstrahlung. The smooth QPO
frequency variations during the low-hard state can be attributed
to a similar increase in the mass accretion rate. This is in
agreement with the QPO production process mentioned above. In this
model the spectrum becomes harder at lower luminosity. This 
fact also has been observed in the ASM data showing a harder
spectrum during the low state (\cite{brad:96}). For a 33 M$_\odot$
black hole (for a stable 67 Hz QPO observed with PCA to represent
the Keplerian rotation of the innermost stable disk around a
non-rotating black hole) and 0.7 Hz QPO, the average shock
front is at a radius of around 100 r$_g$.

\section{Conclusion}

In our 4 days of observations of the galactic superluminal source
GRS 1915+105 in the energy range 2-18 keV, we have observed intensity
variations over time scale of 100 ms to few seconds. The hard X-rays are
found to have time lag compared to the soft X-rays supporting the
Compton up-scattering models of hard X-ray production in black-hole
candidates. Strong QPOs in a narrow range of 0.62 Hz to 0.82 Hz are
from a narrow region in the disk which is stable during the quiescent
state of this source. We show that the observed QPO frequency
variations can be explained if the QPOs are generated by the
oscillations of the shock front in the accretion disk.

{\it Acknowledgements:}
We gratefully acknowledge the support given by Shri K. Thyagrajan,
the Project Director of IRS-P3 satellite, Shri R. N. Tyagi, Manager
IRS programme, Shri R. Aravamudan, Director, ISAC and other
technical and engineering staff of ISAC in making the IXAE project
a success. We particularly thank M. R. Shah, J. P. Malkar,
K. Mukerjee, D. K. Dedhia and P. Shah of our group and N. Upadhyaya,
M. R. Sharma, N. S, Murthy, C. N. Umapathy and L. Abraham of
Technical Physics Devision, ISAC, and other engineering, scientific
and technical staff of other dvisions of ISAC who were involved in
the IRS-P3 project and ISRO tracking facility, for their efforts
in making this experiment successful.

\eject

\begin{figure*}[t]
\vskip 9 cm
\includegraphics{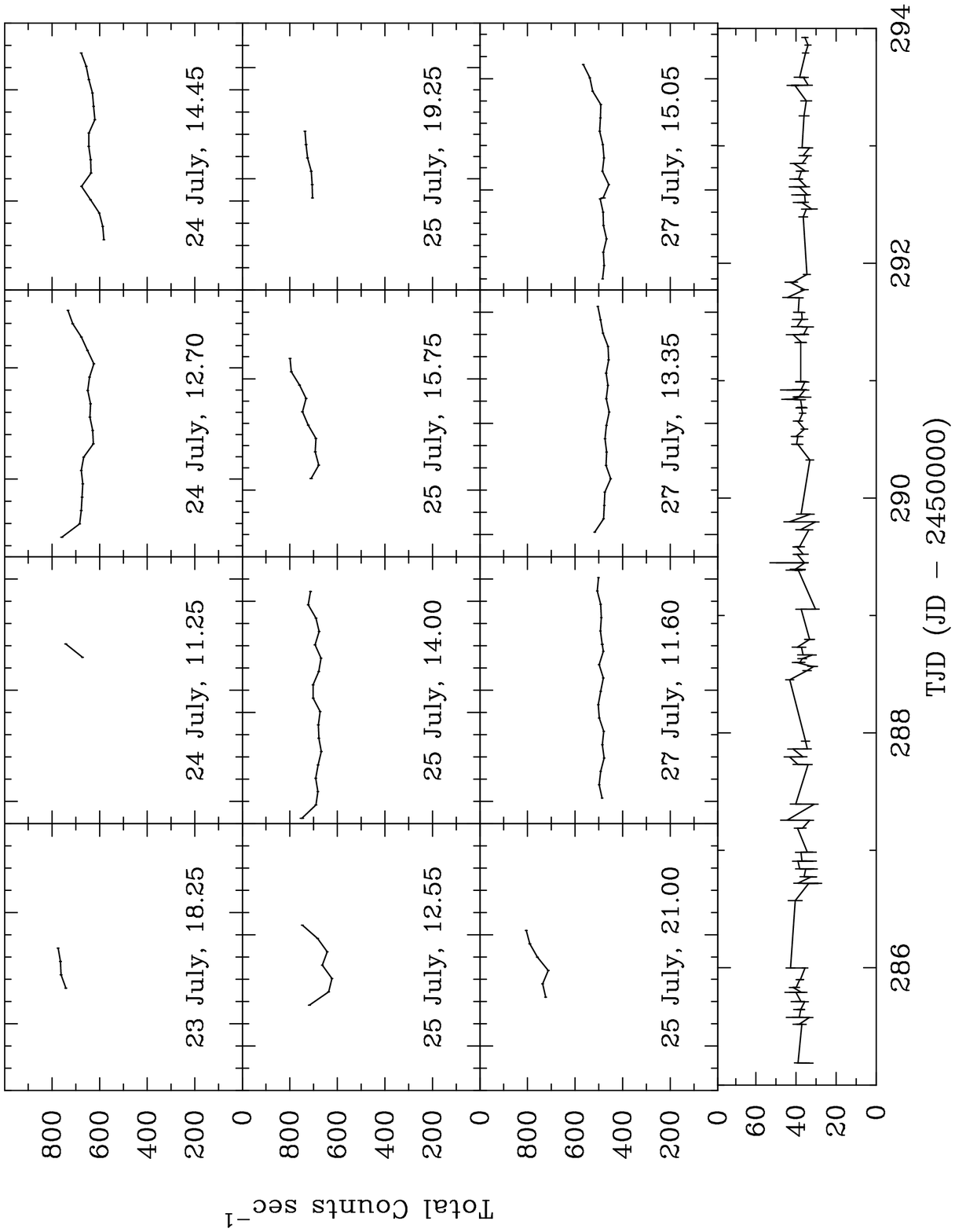} 
\caption{ The light curve of GRS 1915+105 observed with the PPCs. Date
and times of observations are shown in the individual sections.
Each section is for 20 minutes duration and data points of bin size
of 1 minute are plotted. The bottom panel shows the ASM light curve
during July 20-29, each data point is about 90 seconds of observations
with about 5-10 observations every day.}
\end{figure*} 

\begin{figure}[htbp]
\vskip 7.5cm
\includegraphics{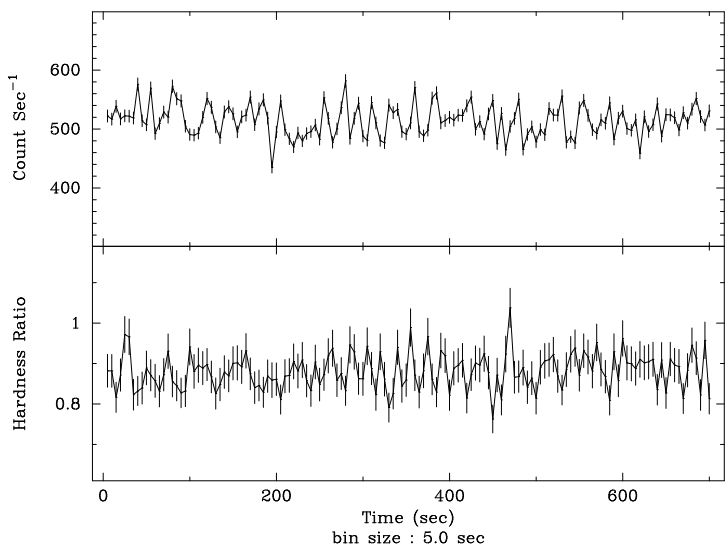} 
\caption{ The intensity and hardness ratio are plotted for one of the PPC
observation slots. Data bin size is chosen to be 5 seconds to reduce the
statistical errors.}
\end{figure} 

\begin{figure*}[htbp]
\vskip 18cm
\includegraphics{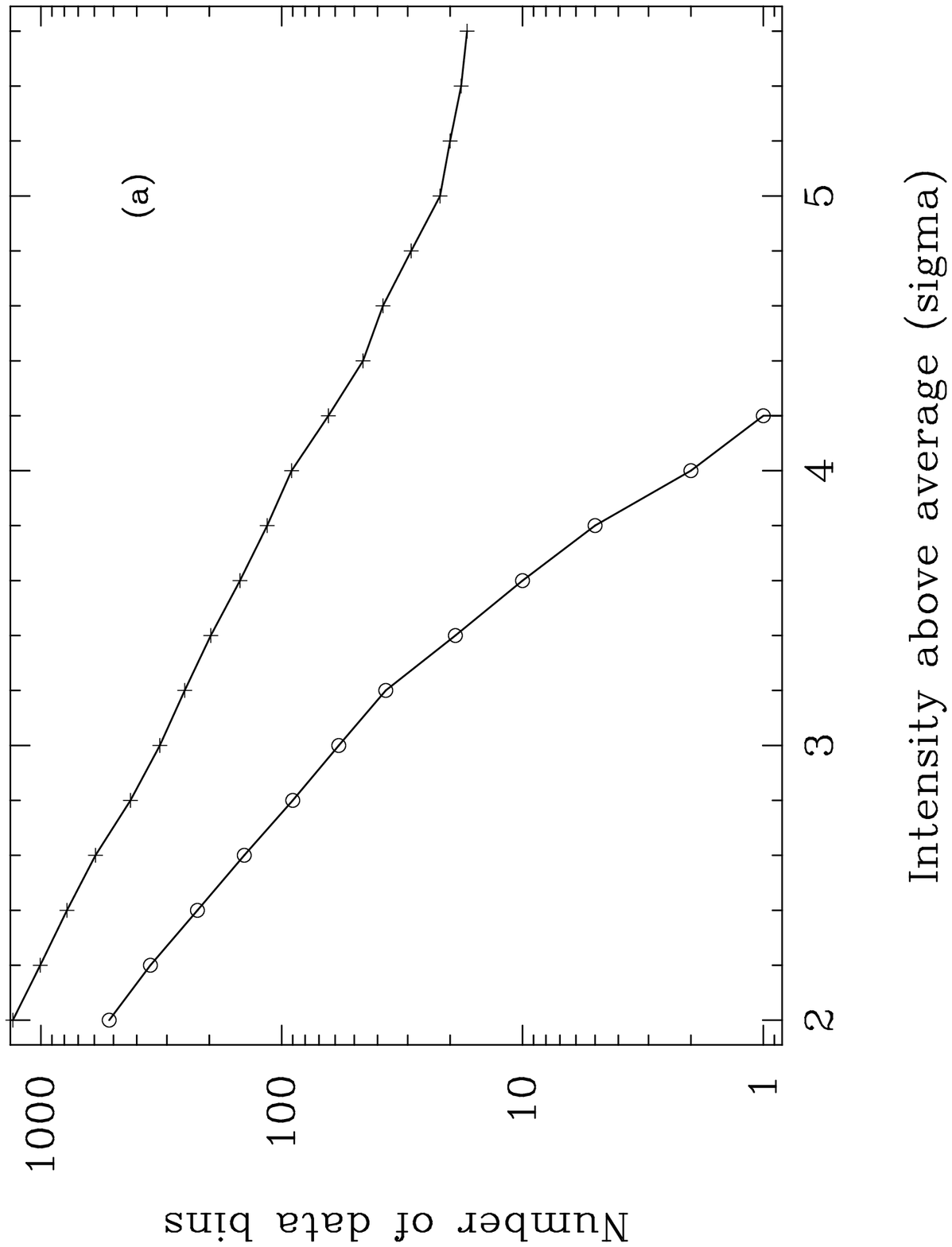} 
\includegraphics{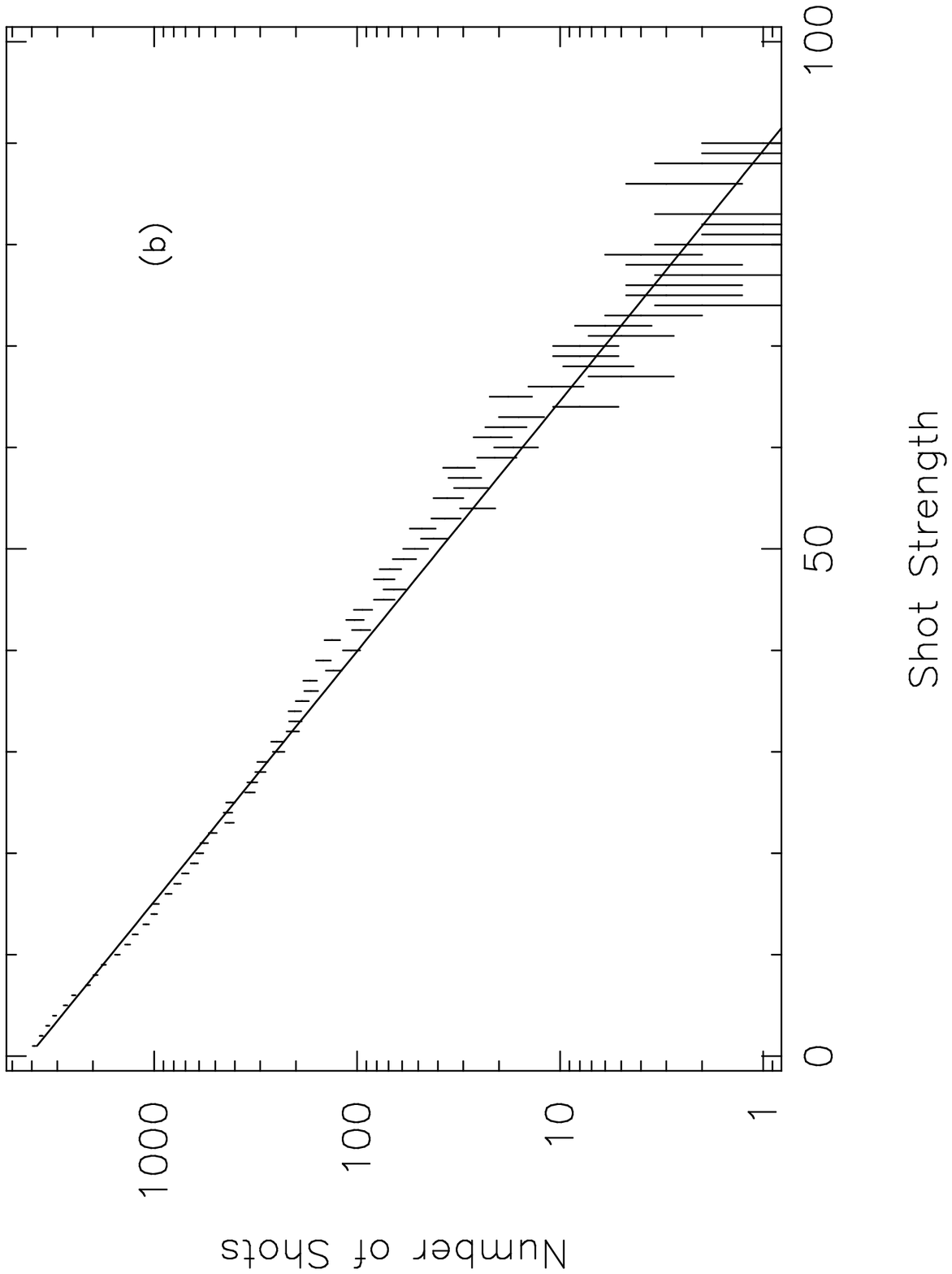} 
\includegraphics{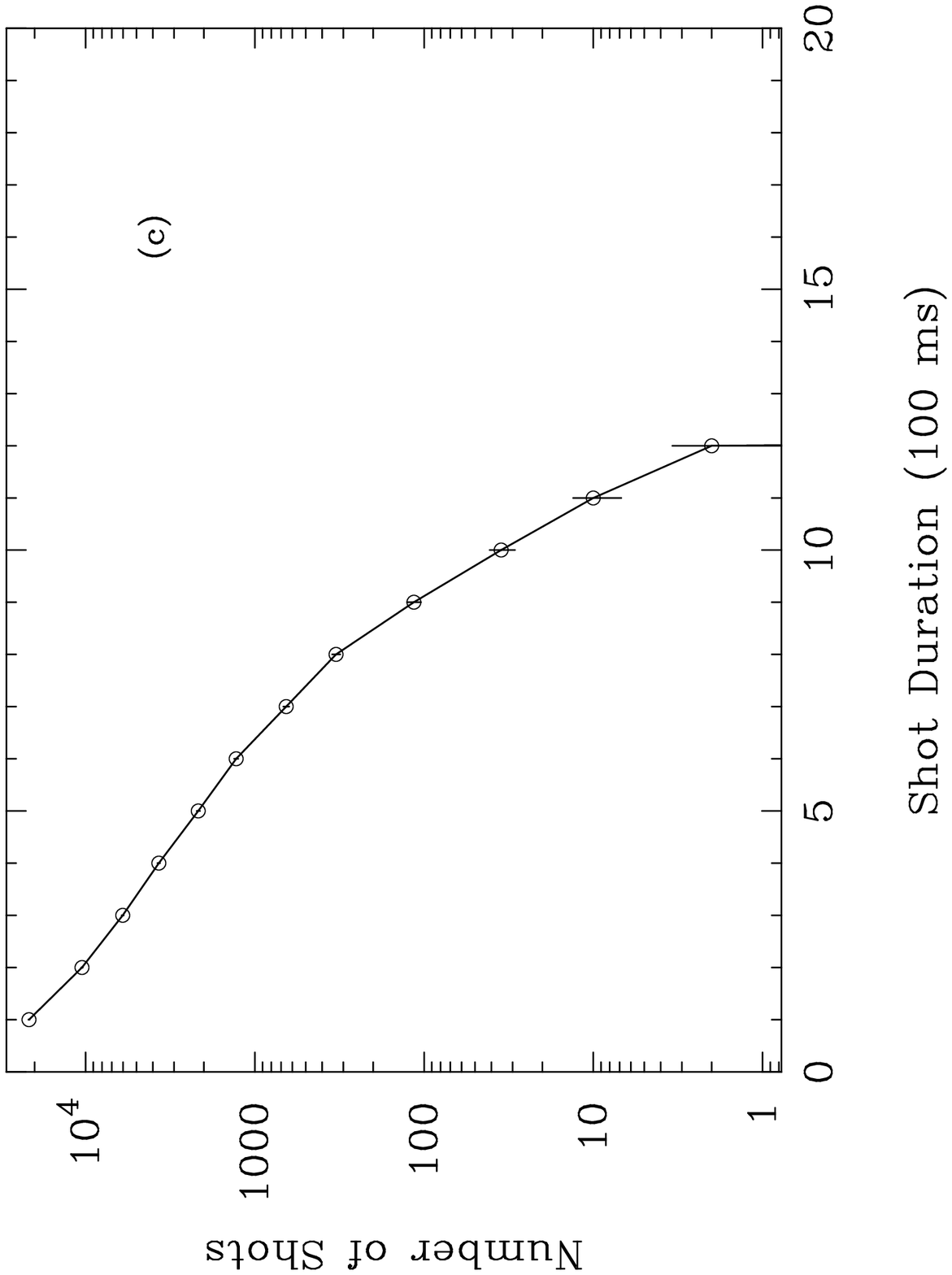} 
\caption{A statistical representation of the intensity variations in GRS 1915+105
observed with the PPCs. The data sets of the three detectors are treated as
independent observations and the results are added.
a). The number of 100 ms data bins exceeding a running
average of 21 data points is plotted as a function of the excess expressed in
$\sigma$. The lower curve is the expected distribution calculated with a 
synthetic light curve with the same running average as in the observed light
curve,
b). the number distribution of the shots as a function of number of
photons in the shots and
c). the number distribution of shots as a function
of shot duration. }
\end{figure*} 

\begin{figure}[htbp]
\vskip 6.0cm
\includegraphics{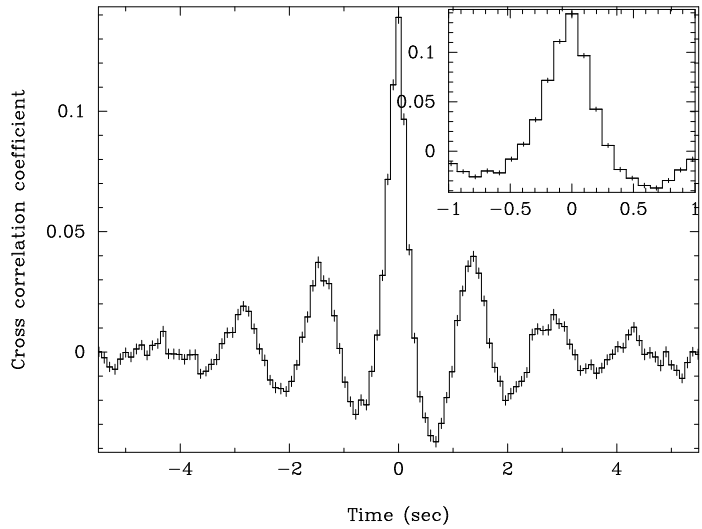} 
\caption{ The cross correlation function for different delays between the
soft X-rays (2-6 keV) and hard X-rays (6-18 keV) is plotted here. The
oscillations in the plot is due to the QPOs at a period of 1.4 seconds.
In the inset is shown the cross correlation function near zero,
asymmetry on the two sides of zero is visible.}
\end{figure} 

\begin{figure}[htbp]
\vskip 6.0cm
\includegraphics{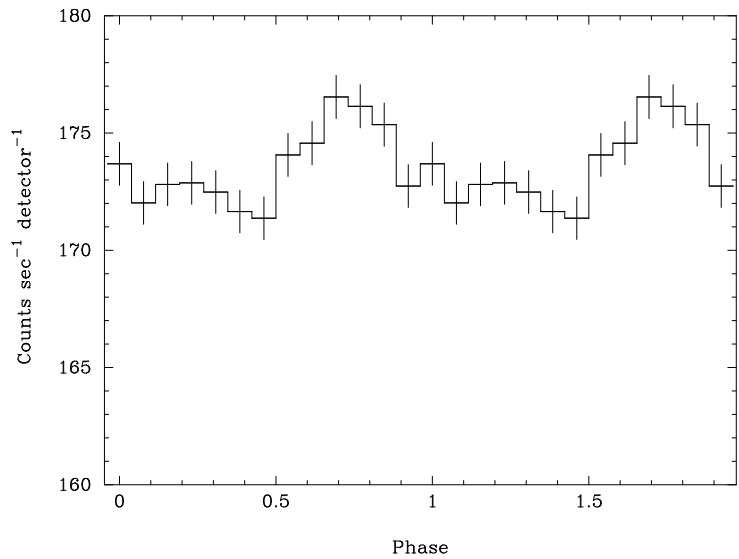} 
\caption{ The pulse profile of GRS 1915+105 folded with the mean
QPO period.}
\end{figure} 

\begin{figure}[htbp]
\vskip 6.0cm
\includegraphics{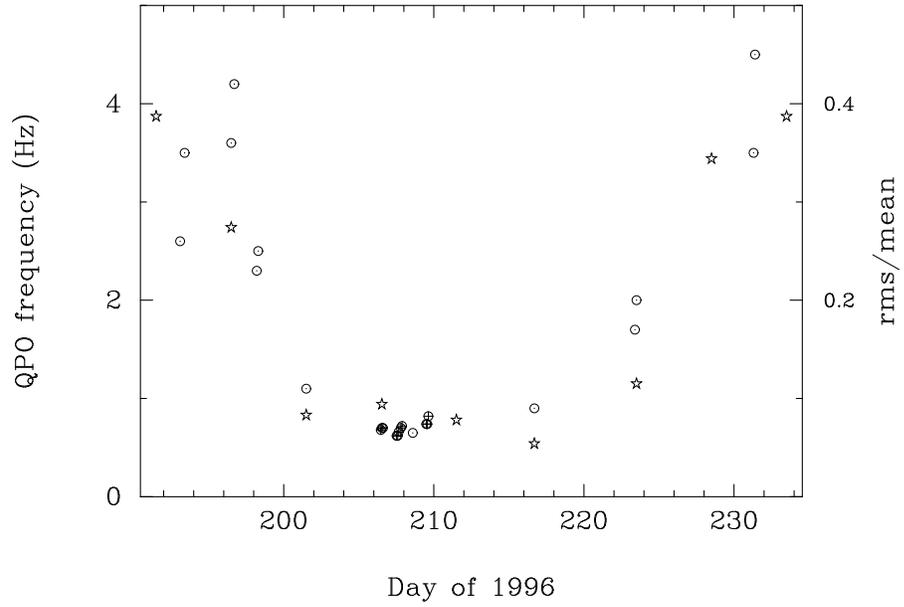} 
\caption{ The QPO frequency (left scale) and rms variation (right scale)
history of GRS 1915+105
during the low-hard state. The crossed circles are PPC observations and
dotted circles are PCA observations. The stars represent the rms/mean during
a 6 days period in the ASM light curve around each point. }
\end{figure} 

\end{document}